\documentclass{vldb}
\usepackage{graphicx}
\usepackage{balance}  % for  \balance command ON LAST PAGE  (only there!)

\usepackage{caption}
\usepackage{amssymb} % main package.
\usepackage{amsmath} % for writing math in between text
\usepackage{mdframed} % for the algorithms frame boxes
\usepackage{comment} % allowing comments in the file
\usepackage{colortbl} % allowing  colors in the tables cells
\usepackage{xcolor} % for mdframe color

\newcommand{\intersection}{\mbox{$\cap$}}

\vldbTitle{NN-based Transformation of Any SQL Cardinality Estimator for Handling DISTINCT, AND, OR and NOT}
\vldbAuthors{Rojeh Hayek, Oded Shmuile}
\vldbDOI{https://doi.org/10.14778/xxxxxxx.xxxxxxx}
\vldbVolume{12}
\vldbNumber{xxx}
\vldbYear{2020}

\begin{document}
\nocite{*} % to display all the uncited references

% ****************** TITLE ****************************************

\title{NN-based Transformation of Any SQL Cardinality Estimator for Handling DISTINCT, AND, OR and NOT}

% possible, but not really needed or used for PVLDB:
%\subtitle{[Extended Abstract]
%\titlenote{A full version of this paper is available as\textit{Author's Guide to Preparing ACM SIG Proceedings Using \LaTeX$2_\epsilon$\ and BibTeX} at \texttt{www.acm.org/eaddress.htm}}}

% ****************** AUTHORS **************************************

\numberofauthors{2} 

\author{
% 1st. author
\alignauthor
Rojeh Hayek\\
       \affaddr{CS Department, Technion}\\
       \affaddr{Haifa 3200003, Israel}\\
       \email{srojeh@cs.technion.ac.il}
% 2nd. author
\alignauthor
Oded Shmueli\\
       \affaddr{CS Department, Technion}\\
       \affaddr{Haifa 3200003, Israel}\\
       \email{oshmu@cs.technion.ac.il}
}

\maketitle

\begin{abstract}
SQL queries, with the AND, OR, and NOT operators, constitute a broad class of highly used queries. Thus, their cardinality estimation is important for query optimization. In addition, a query planner requires the set-theoretic cardinality (i.e., without duplicates) for queries with DISTINCT as well as in planning; for example, when considering sorting options. Yet, despite the importance of estimating query cardinalities in the presence of DISTINCT, AND, OR, and NOT, many cardinality estimation methods are limited to estimating cardinalities of only conjunctive queries with duplicates counted.

The focus of this work is on two methods for handling this deficiency that can be applied to \textit{any} limited cardinality estimation model. First, we describe a specialized deep learning scheme, PUNQ, which is tailored to representing conjunctive SQL queries and predicting the percentage of unique rows in the query’s result with duplicate rows. Using the predicted percentages obtained via PUNQ, we are able to transform \textit{any} cardinality estimation method that only estimates for conjunctive queries, and which estimates cardinalities with duplicates (e.g., MSCN), to a method that estimates queries cardinalities without duplicates. This enables estimating cardinalities of queries with the DISTINCT keyword. In addition, we describe a recursive algorithm, GenCrd, for extending \textit{any} cardinality estimation method $M$ that only handles conjunctive queries to one that estimates cardinalities for more general queries (that include the AND, OR, and NOT operators), without changing the method $M$ itself.

Our evaluation is carried out on a challenging, real-world database with general queries that include either the DISTINCT keyword or the AND, OR, and NOT operators. The methods we use are obtained by transforming different basic cardinality estimation methods. The evaluation shows that the proposed methods obtain accurate cardinality estimates with the same level of accuracy as that of the original transformed methods.
\end{abstract}

\section{Introduction}
Recently, several neural network (NN) based models were proposed for solving the cardinality estimation problem, such as the multi-set convolutional network (MSCN) model of \cite{MSCN}, and the containment rate network (CRN) model of \cite{CRN}. 

\cite{MSCN} introduced the MSCN model, a sophisticated NN tailored for representing SQL queries and predicting cardinalities, resulting in improved cardinality estimates as compared with state-of-the-art methods. 

In \cite{CRN} we introduced a method, Cnt2Crd($M$), for estimating cardinality based on using any model $M$ for estimating the percentage (rate) of containment between query results of two input queries. To estimate the containment rates between pairs of queries, \cite{CRN} introduced the CRN (NN-based) model.
In addition, \cite{CRN} proposed a simple transformation that converts any cardinality estimation model $M$ (e.g., MSCN) into one for estimating containment rates $M'$. As a result, Cnt2Crd($M'$) obtains more accurate cardinality estimates than the original model ($M$). For this, Cnt2Crd($M'$) is referred to as Improved $M$ (e.g., Improved MSCN).
Using these techniques, \cite{CRN} showed that Cnt2Crd(CRN), referred to henceforth simply as CRN, produced superior predictions in most settings, compared to Improved MSCN, Improved postgreSQL, MSCN, and postgreSQL.

\begin{comment}
To use Cnt2Crd, a model for estimating cardinality may be first converted into one for estimating containment rates. For example, MSCN and postgreSQL were converted to models MSCN’ and PostgreSQL’ to estimate containment rates using a simple transformation \cite{CRN}. The resulting cardinality estimation models Cnt2Crd(MSCN’) and Cnt2Crd(PostgreSQL’) models are referred to as Improved MSCN, and Improved PostgreSQL, respectively.
In addition, [9] introduced the CRN model for estimating containment rates directly. 
Cnt2Crd(CRN), referred to henceforth simply as CRN, produced superior predictions in most settings, compared to Improved MSCN, Improved postgreSQL, MSCN, and postgreSQL.

It is important to emphasize that, alongside the improvement by the models that adapt the Cnt2Crd approach, there is a disadvantage in that these models run time is slower than that of MSCN and PostgreSQL and consumes additional space.
\end{comment}

Both MSCN and CRN are superior in estimating cardinalities compared to basic cardinality estimation methods such as PostgreSQL \cite{postgreSQL}, Index Based Join Sampling (IBJS) \cite{IBJS} and Random Sampling (RS) \cite{RS1,RS2}. However, both MSCN and CRN, and other recently developed models, are limited to supporting conjunctive queries only, ignoring the fact that queries with the AND, OR, and NOT operators constitute a broad class of frequently used queries.
Furthermore, the abovementioned models were designed to only estimate cardinalities \textit{with} duplicates, thus further limiting the class of supported queries as queries with the DISTINCT keyword are simply not supported.
In addition to not supporting queries with the DISTINCT keyword, for various intermediate results, a query planner requires the set-theoretic cardinality (without duplicates) even when the DISTINCT keyword is not used explicitly in the query. For example, in employing counting techniques for handling duplicates, in considering sorting options, and in creating an index or a hash table.

With all the improvements of the recently proposed models, the abovementioned limitations pose a significant problem to the applicability of these models. In real world systems, supporting (more general) queries with the AND, OR, and NOT operators and the DISTINCT keyword is essential.
The limited models\footnote{In this paper, the term "limited models" refers to models that support only conjunctive queries whose estimates count duplicates.}, lacking AND, OR, NOT and DISTINCT support, vary significantly. Thus, restructuring each one, separately, to support more general queries, is complex and potentially error-ridden. 

Estimating cardinalities without duplicates (e.g., queries with the DISTINCT keyword) or supporting queries with the AND, OR, and NOT operators using specialized models is indeed interesting. 
However, many existing limited cardinality estimation models, including state of the art models (e.g., MSCN, CRN), have been already developed and were shown to be very accurate in estimating conjunctive queries' cardinalities. This points to a different direction. Instead of trying to build, from scratch, specialized models to handle more general queries\footnote{In this paper, the term "general queries" refers, depending on the context, to queries that include: the DISTINCT keyword and/or the AND, OR and NOT operators.}, we propose two different methods that utilize already developed efficient limited models to handle more general queries while \textit{retaining} the original models quality of estimates.

That is, the thrust of this paper is using two methods for simply and efficiently extend \textit{any} limited model to handle queries that were not supported by that original model. Additionally we illustrate how these methods can extend any limited model while \textit{retaining} the original model’s quality. Therefore, the experiments we exhibit aim to illustrate this fact, and are not intended to show which model is more accurate following the extension, as the original examined limited models, per se, are not the focus of this paper (they were previously developed). The examined limited models were simply used as \textit{examples} for our extension methods.

To tackle the lack of support for the set-theoretic cardinality estimation (i.e., cardinalities without counting duplicates), we use a specialized deep learning method, PUNQ. The PUNQ model is tailored to representing conjunctive SQL queries and predicting the percentage of unique rows within the query’s result with duplicates. Using the predicted percentages, we propose a simple technique to extend \textit{any} limited model to estimate set-theoretic cardinalities.

\begin{comment}
Given a query $Q$ whose set-theoretic cardinality needs to be estimated, and a limited model $M$, the set-theoretic cardinality can be estimated as follows. Assuming that $U$ is the predicted percentage of unique rows in the query’s result obtained from PUNQ, and $C$ is the estimated cardinality with duplicates, obtained from model $M$, the value $C \cdot U$ is the estimated set-theoretic cardinality.
\end{comment}

The PUNQ model uses a tailored vector representation for representing the input queries, which enables us to express the query’s features. An input query is converted into multiple sets, each representing a different part of the query. Each element, of one of these sets, is represented by a vector. These vectors are then passed into a specialized neural network that combines them into a single representative vector that represents the whole input query. Finally, using another neural network with the representative vector as input, PUNQ predicts the percentage of unique rows in the query’s result with duplicates. Thus, the PUNQ model relies on the ability of its neural network to learn the (representative) vector representation of queries. As a result, we obtain a small and accurate model for estimating the percentage of unique rows in a query result that includes duplicates.

Queries that include the AND, OR, and NOT operators in their WHERE clauses constitute a broad class of frequently used queries. Their expressive power is roughly equivalent to that of the relational algebra. Therefore, estimating cardinalities for such queries is essential. To estimate the result cardinality of such queries, we introduce a recursive algorithm, GenCrd, that estimates the cardinality of a general query (that includes the AND, OR, and NOT operators), using \textit{any} limited model that only estimates cardinalities for conjunctive queries, as a "subroutine". 

We evaluated our extension methods (both the PUNQ model and GenCrd algorithm), on a challenging, real-world IMDb database \cite{HowGoodCar} with general queries that either include the DISTINCT keyword or the AND, OR, and NOT operators. In addition, the examined queries include join crossing correlations; these are known to present a special challenge to cardinality estimation methods \cite{HowGoodCar,crdHard,JoinCross}.

The extension methods were applied to four limited models that were used as \textit{examples}, MSCN \cite{MSCN}, CRN \cite{CRN}, Improved MSCN \cite{CRN}, and Improved PostgreSQL \cite{CRN}. The extended methods were shown to produce cardinality estimates with the same level of accuracy as that of the original methods. This indicates the practicality of our extension methods, offering a simple and efficient tool for extending \textit{any} limited cardinality estimation method, to support queries that are more general. As a sanity check, we also compare the four extended methods with PostgreSQL \cite{postgreSQL}, as PostgreSQL supports all kinds of queries without using our extension methods.

This paper has four main contributions:
\begin{enumerate}
    \item We introduced two different extension methods,\\PUNQ($M$) and GenCrd($M$), which can extend \textit{any} limited model $M$ to supporting more general queries.
    \item Although limited models implementation vary, the proposed extension methods can be applied to \textit{any} limited model regardless its implementation.
    \item Some of the limited models are considered state-of-the-art in terms of estimation accuracy. Our extension methods (on the examined models) proved to retain the original limited model's quality of estimates, thus presenting a simple and efficient blueprint for expanding the limited models class of supported queries.
    \item Using the proposed extension methods, we factor out the code needed to expand each limited model, resulting in an improved software architecture.
\end{enumerate}

The rest of this paper is organized as follows. In Sections \ref{Supporting DISTINCT keyword} and \ref{PUNQ Evaluation}, we describe and evaluate the PUNQ model for extending any limited model to support the DISTINCT keyword (i.e. estimating set-theoretic cardinalities). In Sections \ref{Supporting AND, OR, NOT operators} and  \ref{GenCrd Evaluation}, we describe and evaluate recursive algorithm GenCrd, that extends any limited model to support queries with the AND, OR, and NOT operators. Finally, in Section \ref{Related Work}, we present related work, and in Section \ref{Conclusion} we present our conclusions, and plans for future work.

\newpage
\section{Supporting the DISTINCT keyword}
\label{Supporting DISTINCT keyword}
Several cardinality estimation models are designed to estimate cardinalities with duplicates, ignoring the importance of the set-theoretic cardinality estimation in query optimization. We would like to extend such models, \textit{without} changing them, to produce set-theoretic cardinality estimates. To this end, we design a NN based model, PUNQ, which is tailored to estimate the \textit{uniqueness rate} of its input query (see Section \ref{Uniqueness Rate Definition}).
We propose this technique as it can be applied to \textit{any} limited model (regardless the limited model's implementation). Thus, using the PUNQ model, we can transform \textit{any} limited model to estimate cardinalities without duplicates (thereby supporting queries with the DISTINCT keyword).

\subsection{Uniqueness Rate}
\label{Uniqueness Rate Definition}
We define the uniqueness rate of query $Q$ on a \textit{specific} database $D$, as follows. \textit{The uniqueness rate of query $Q$ equals to $x\%$ on database $D$ if precisely $x\%$ of $Q$'s execution result rows (with duplicates) on database $D$ are unique}. The uniqueness rate is formally a function from \textbf{QxD} to \textbf{[0,1]}, where \textbf{Q} is the set of all queries, and \textbf{D} of all databases. This function can be directly calculated using the cardinality of the results of query $Q$ as follows:
$$ x\% = \frac{||Q(D)||}{|Q(D)|} $$
where, $Q(D)$ denotes $Q$'s execution result on database $D$. The operator $||\cdot||$ returns the set theoretic cardinality, and the operator $|\cdot|$ returns the cardinality with duplicates. (In case $Q$'s execution result is empty, then $Q$ has 0\% unique rows). 
For clarity, we do not mention the \textit{specific} database $D$, that it is usually clear from context.

For example, suppose that query $Q$'s one column result, on database $D$, includes 10 rows with value 1, and 4 rows with value 2 and 1 row with value 3. By definition, query $Q$'s cardinality with duplicates equals to 15, and the cardinality without duplicates equals to 3. Thus, $Q$'s uniqueness rate equals to 20\%. This is true, since by definition:
$$ Q's\ Uniqueness\ Rate = \frac{||Q(D)||}{|Q(D)|} = \frac{3}{15} = 0.2$$

Suppose we are given a cardinality estimate $C$ with duplicates of some query $Q$ on $D$, obtained from \textit{some} cardinality estimation model. By uniqueness rate definition, we can easily obtain the set-theoretic cardinality estimation simply by multiplying the cardinality estimate $C$ with the uniqueness rate of query $Q$. Therefore, using the uniqueness rates of queries we can simply transform \textit{any} limited cardinality estimation method to one that estimates set-theoretic cardinality. Hence, in the following sections we aim to estimate accurately and efficiently the uniqueness rates of queries. For this, we use the subsequently described PUNQ model.

\subsection{Learned Uniqueness Rates}
\label{Learned Uniqueness Rates}
In applying machine learning to the uniqueness-rate estimation problem, we face the following common issues that are faced by any other deep learning technique. 
\begin{itemize}
    \item Cold start problem: obtaining the initial training dataset. 
    \item Determining which supervised model should be used.
    \item Featurization: determining useful representations of the inputs and the outputs of the model.
\end{itemize}
The ways these issues are addressed determine the success of the machine learning based  model (PUNQ in our case). In the following sections we describe how each issue is addressed.

\subsection{Cold Start Problem}
\label{Defining the database and the development set}
\subsubsection{Defining the Database}
\label{Defining the database}
We generated a training-set of queries, and later on evaluated our model on it, using the IMDb database. The IMDb database stores information about over 2.5M movies. There are about 4M related movie actors, directors, and production companies. IMDb data has many internal correlations. It is thus challenging for cardinality estimators \cite{HowGoodCar}. As IMDb was previously used in \cite{MSCN,CRN} by the methods we extend, we also use this database to evaluate our PUNQ based method.

\subsubsection{Generating the Development Dataset}
\label{queries generator} This is the "cold start problem". We obtain an initial training corpus using a specialized queries generator. The generator randomly generates queries based on the IMDB schema. It uses potential columns' values for constants. The queries generator is constructed similarly to the ones used in \cite{MSCN,CRN}.
The queries generator forms a query as follows. The number of tables in the query's FROM clause is chosen randomly, The specific tables are also chosen randomly. Join edges are then randomly set. Next, for each chosen table $t$, column predicates (comparison to constants) are chosen. The generator first chooses the number $p_t$ of columns predicates on table $t$. Then, it generates $p_t$ random column predicates. A column predicate is defined by randomly chosen three values: a column name $Col$, a type $(<,\ =,\ or\ >)$, and a constant within the range of values in the actual database column $Col$.

Observe that the specifics of the SELECT clause have \textit{no impact} on the result cardinality when considering cardinalities with duplicates (i.e., no DISTINCT keyword). 
Since the MSCN model \cite{MSCN} and the CRN model \cite{CRN} were designed to estimate cardinalities with duplicates, the queries generator used by \cite{CRN,MSCN} can be thought of as assuming a SELECT * clause for all the generated queries. However, the specifics of the SELECT clause are \textit{extremely important} for our task (predicting uniqueness rates). Therefore, we reconfigure the queries generator with the following additional step. After generating the query with the SELECT * clause, we replace the * in the SELECT clause by randomly choosing the number of columns $c$ to be in the query’s SELECT clause, and then choosing at random $c$ different columns from the possible columns of the chosen tables.

In order to avoid a combinatorial explosion of the number of possible queries, and to facilitate the learning of the model, we create queries with up to two joins only. We also examine how the learned models, learned over queries with this small number of joins, generalize to a larger number of joins.

Once generating the dataset of queries, we execute the dataset queries on the IMDb database. This allows us to obtain their true uniqueness rates (ground truth). In this process, we obtain an initial training set for our model. It consists of 20,000 queries with zero to two joins. We split the training samples into 80\% training samples and 20\% validation samples.

\subsection{Model}
\label{Model}
We form the PUNQ model for estimating the uniqueness rates (the percentage of rows in the input query’s result with duplicates that are unique). PUNQ was inspired by the architectures of the MSCN model \cite{MSCN} and the CRN \cite{CRN} model. The PUNQ model has three main stages. In the first stage, it converts the input query into a set $V$ of vectors. A vector represents either the query's columns, tables, joins or columns predicates. In stage two, using a specialized neural network, $MLP_{mid}$, the set $V$ of vectors is transformed into $Qvec$, a single representative vector that represents the \textit{whole} input query. In stage three, the percentage of unique rows is estimated using $MLP_{out}$, a two-layers neural network.

\label{First Stage}
\subsubsection{First Stage, from $Q$ to $V$}
We represent each query $Q$ as a collection of four sets $(A,T,J,P)$. 
\begin{itemize}
    \item $A$: set of the attributes in $Q$'s SELECT clause.
    \item $T$: set of the tables in $Q$'s FROM clause.
    \item $J$: set of the join clauses in $Q$'s WHERE clause. 
    \item $P$: set of the columns predicates in $Q$'s WHERE clause.
\end{itemize}
Each element in each of the 4 sets is then represented with a vector. Together these vectors form the set $V$ of vectors. To facilitate learning, all the vectors in $V$ are of the same dimension and the same format as depicted in Table \ref{table:Vector Segmentation}, where:
\begin{itemize}
    \item $\#T$: the number of tables in the whole database.
    \item $\#C$: the number of columns in the whole database.
    \item $\#O$: the number of all possible operators in predicates.
\end{itemize}
Hence, the vector size is $\#T + 4*\#C + \#O + 1$, denoted as $L$.

\begin{table}[h!]
\centering
\resizebox{\linewidth}{!}{
\begin{tabular}{c|c|c|c|c|c|c|c}
\hline
\hline
$\bold{Type}$ & $\bold{Att.}$ & $\bold{Table}$ & \multicolumn{2}{c|}{$\bold{Join}$} & \multicolumn{3}{c}{$\bold{Column\ Predicate}$} \\
\hline
$\bold {Segment}$ & \textbf{A-seg} & \textbf{T-seg} & \textbf{J1-seg} & \textbf{J2-seg} &  \textbf{C-seg} & \textbf{O-seg} & \textbf{V-seg} \\
\hline
$\bold {Seg.\ size}$ & $\#C$ & $\#T$ & $\#C$ &  $\#C$ &  $\#C$ & $\#O$ & $1$ \\
\hline
$\bold {Feat.}$ & $1hot\ vec.$ & $1hot\ vec.$ & $1hot\ vec.$ & $1hot\ vec.$ &  $1hot\ vec.$ & $1hot\ vec.$ & $norm.$ \\
\hline
\hline
\end{tabular}
}
\caption{Vector Segmentation.}
\label{table:Vector Segmentation}
\end{table}

\begin{figure*}[h!]
\begin{center}
  \includegraphics[width=\linewidth]{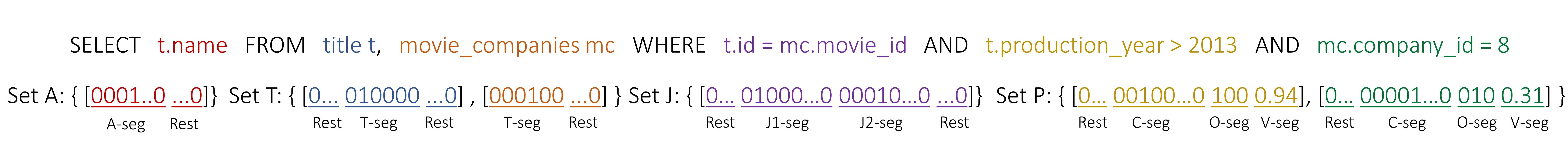}
  \caption{Query featurization as sets of feature vectors obtained from sets $A$, $T$, $J$ and $P$ (Rest denotes zeroed portions of a vector).}
  \label{example}
\end{center}
\end{figure*}

For each element of the $A,T,P,$ and $J$ sets, we create a vector of dimension $L$, that includes zeros in all its segments, except for those that are relevant for representing its element, as follows. (see a simple example in Figure \ref{example})

Elements of set $A$, are represented in the A-seg segment. Elements of set $T$, are represented in the T-seg segment.
Elements $(col1,=,col2)$ of set $J$ are represented in two segments. $col1$ and $col2$ are represented in J1-seg and J2-seg segments, respectively.
Elements $(col,op,val)$ of set $P$ are represented in three segments. $col$ and $op$ are represented in the C-seg and O-seg segments, respectively; $val$ is represented in the V-seg segment.

All the segments, except for the V-seg, are represented by a binary vector with a single non-zero entry, uniquely identifying the corresponding element (one-hot vector). For example, T-seg segment, is represented with one-hot vector uniquely identifying the corresponding table name.

In the V-seg segment, we represent the predicates values. Since their domain is usually large, representing them using one hot vectors is not practical. Therefore, these values are represented using their normalized value (a value $\in [0, 1]$). 

\begin{comment}
\begin{itemize}
    \item Each attribute $c \in C$ is represented by a unique one-hot vector placed in the A-seg segment. 
    \item Each table $t \in T$ is represented by a unique one-hot vector placed in the T-seg segment.
    \item Each join clause of the form $(col1,=,col2) \in J$ is represented as follows. $col1$ and $col2$ are represented by a unique one-hot vectors placed in J1-seg and J2-seg segments, respectively.
    \item Each predicate of the form $(col,op,val) \in P$ is represented as follows. $col$ and $op$ are represented by a unique one-hot vectors placed in the C-seg and O-seg segments, respectively. $val$ is represented as a normalized value $\in [0, 1]$, normalized using the minimum and maximum values of the respective column, placed in the V-seg segment.
\end{itemize}
\end{comment}

\subsubsection{Second Stage, from $V$ to $Qvec$}
Given the set of vectors $V$, we present each vector of $V$ as input to $MLP_{mid}$, a fully-connected one-layer neural network. $MLP_{mid}$ transforms each presented vector into a vector of dimension $H$. The final representation for the whole of set $V$ is then given by $Qvec$, the average over the individual transformed representations of its elements, i.e.,
$$ Qvec = \frac{1}{|V|} \sum_{v \in V} MLP_{mid} (v)$$
$$ MLP_{mid}(v) = RelU(vU_{mid} + b_{mid})$$
where $U_{mid} \in R^{LxH}$, $b_{mid} \in R^{H}$ are the learned weights and bias, and $v \in R^{L}$ is the input row vector. We choose an average (instead of, say, sum) to facilitate generalization to a different number of elements in $V$.  Had we used sum, the overall magnitude of $Qvec$ would depend on the cardinality of the set V.

\subsubsection{Third Stage, from $Qvec$ to $Uniqueness\ rate$}

Given the representative vector of the input query, $Qvec$, we aim to predict the uniqueness rate. To do so, we use $MLP_{out}$, a fully-connected two-layer neural network, to compute the estimated uniqueness rate of the input query. 

$MLP_{out}$ takes as input the $Qvec$ vector of size $H$. The first layer in $MLP_{out}$ converts the input vector into a vector of size $0.5H$. The second layer converts the obtained  vector of size $0.5H$, into a single value representing the uniqueness rate. This architecture allows $MLP_{out}$ to learn the uniqueness rate function.

$$ \hat{y} = MLP_{out}(Qvec)$$
$$ MLP_{out}(v) = Sigmoid(ReLU(vU_{out1} + b_{out1})U_{out2} + b_{out2}) $$
where $\hat{y}$ is the estimated uniqueness rate, $U_{out1} \in R^{Hx0.5H}$,  $b_{out1} \in R^{0.5H}$ and $U_{out2} \in R^{0.5Hx1}$,  $b_{out2} \in R^{1}$ are the learned weights and bias.

We use the empirically strong and fast to evaluate $ReLU$\footnote{ReLU(x) = max(0,x); see \cite{ActivationFunc}.} activation function for hidden layers units in all our neural networks.
The uniqueness rate values are in the range [0,1]. Therefore, in the final step, we apply the $Sigmoid$\footnote{Sigmoid(x) = $1/(1+e^{-x})$; see \cite{ActivationFunc}.} activation function in the second layer to output a float value in this range. It follows that we do not apply any featurization on the uniqueness rates (the output of the model) and the model is trained with the actual uniqueness rate values without any intermediate featurization steps.

\subsection{Training and Testing Interface}
The PUNQ model building includes two main steps, performed on an immutable snapshot of the database. We first generate a random training set as described in Section \ref{Defining the database and the development set}. Second, we train the PUNQ model, using the training data, until the mean q-error of the validation test starts to converge to its best absolute value. In the training phase, we use the early stopping technique \cite{earlyStopping}.

We train the PUNQ model to minimize the mean q-error \cite{qerror}. The q-error is a widely-used metric in the cardinality estimation problem domain. The q-error is the ratio between an estimate and the actual value.

Formally, assuming that $y$ is the true (actual) uniqueness rate, and $\hat{y}$ the estimated rate, then the q-error is defined as follows.
$$ q-error(y,\hat{y})\ =\ \hat{y} > y\ ?\ \frac{\hat{y}}{y}\ :\ \frac{y}{\hat{y}}$$

Following the training phase, and in order to predict the uniqueness rate of an input query, the (encoded) query is simply presented to the PUNQ model, and the model outputs the estimated uniqueness rate, as described in Section \ref{Model}. We train and test our model using the Tensor-Flow framework \cite{tensorFlow}, using the Adam training optimizer \cite{adam}.

\subsection{Hyperparameter Search \& Model Costs}
\label{Hyperparameter Search}
To optimize our model's performance, we searched the hyperparameter space. We considered different settings, where we varied the number of the batch size (16, 32, 64, ..., 2048), hidden layer size (16, 32, 64, ..., 1024), and learning rate (0.001, 0.01). We examined all the resulting 112 different hyperparameters combinations.
It turned out that the hidden layer size has the most impact on the models accuracy on the validation test. Until it reaches the best result, the bigger the hidden layer size is, the more accurate the model is on the validation test. Afterwards, quality declines due to over-fitting. Further, both the learning rate and the batch size mainly influence the training convergence behavior rather than the model accuracy.

\begin{figure}[h!]
\begin{center}
  \includegraphics[width=\linewidth]{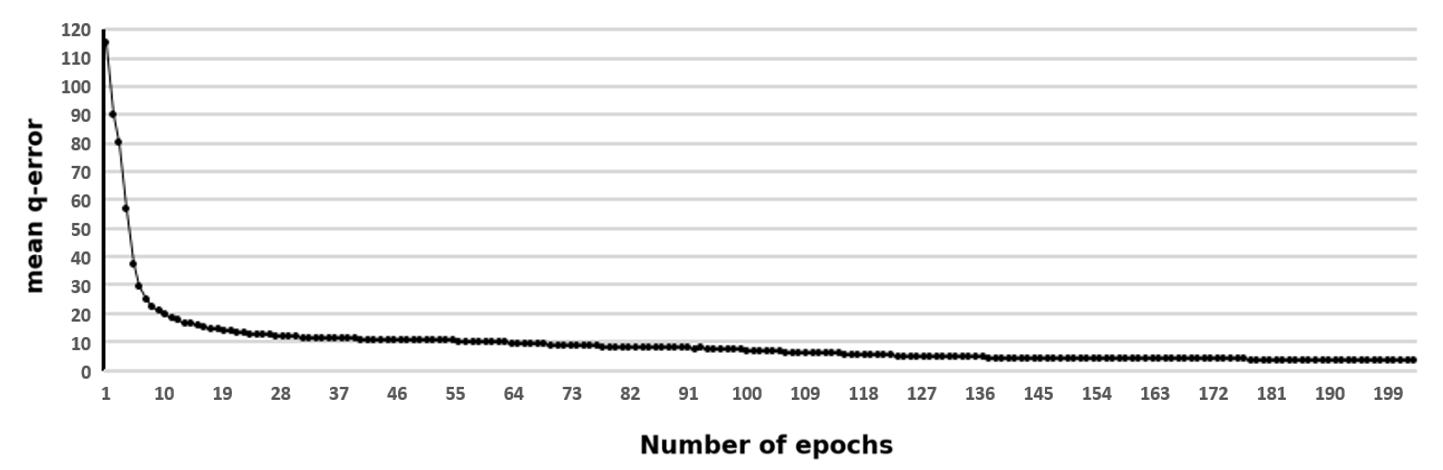}
  \caption{Convergence of the mean q-error.}
  \label{validationConverage}
\end{center}
\end{figure}

Averaged over 5 runs over the validation set, the best configuration has a 128 batch size, a 512 hidden layer size, and a 0.001 learning rate. These settings are thus used throughout our model evaluation. Under these settings, the PUNQ model converges to a mean q-error of approximately 3.5 on the validation set, after running about 200 passes on the training set (see Figure \ref{validationConverage}). Averaged over 5 runs, the 200 epochs training phase takes nearly 30 minutes. The disk-serialized model size is about 0.5MB. This includes all the learned wights and biases as described in Section \ref{Model}.

The prediction time of uniqueness rates, using the PUNQ model, consists of performing all the PUNQ model's stages as described in Section \ref{Model}. On average, the whole process takes 0.05ms, per query. This includes the overhead introduced by the Tensor-Flow framework.

\section{PUNQ Evaluation}
\label{PUNQ Evaluation}
In this section, we examine how well our extension model (PUNQ model), transforms limited cardinality estimation models that predict cardinalities with duplicates, into ones that predict set-theoretic cardinalities. To this end, we extended four different limited models, MSCN, CRN, Improved PostgreSQL, and Improved MSCN \cite{CRN,MSCN}. Each model was extended as detailed below. For clarity, we denote the extended model of model $M$ as PUNQ($M$). Given query Q whose set-theoretic cardinality needs to be estimated (i.e., without duplicates), PUNQ($M$) functions as follows.
\begin{itemize}
    \item calculate $C$ = $M(Q)$, model's $M$ estimated cardinality with duplicates of query $Q$. 
    \item calculate $U$ = $PUNQ(Q)$, i.e., the uniqueness rate of query $Q$, obtained from the PUNQ model.
    \item return $U \cdot C$, the predicted set-theoretic cardinality.
\end{itemize}
We extend each of the four abovementioned limited models and test them over two test workloads. In addition, we compare our results with those of the PostgreSQL version 11 cardinality estimation component \cite{postgreSQL}, as postgreSQL supports all SQL queries. To estimate cardinality without duplicates using PostgreSQL, we added the DISTINCT keyword in the queries' select clauses, and then used the ANALYZE command to obtain the estimated (set-theoretic) cardinalities.

The workloads were created using the same queries generator (using a different random seed) as introduced in Section \ref{queries generator}. In order to ensure a fair comparison between the models, we trained CRN, MSCN and the PUNQ models with the same types of queries, where all the models training sets were generated using similar queries generators. In particular, these models were trained with conjunctive queries with only up to two joins only.

\subsection{Evaluation Workloads}
\label{PUNQ Evaluation Workloads}
The evaluation uses the IMDb dataset, over two different query workloads:
\begin{itemize}
    \item UnqCrd\_test1, a synthetic workload with 450 unique queries, with zero to two joins.
    \item UnqCrd\_test2, a synthetic workload with 450 unique queries, with zero to \textit{five} joins. This dataset is designed to examine how the models generalize beyond 2 joins.
\end{itemize}

\begin{table}[h!]
\centering
\resizebox{\linewidth}{!}{
\begin{tabular}{lccccccc}
\hline
\hline
$\bold {number\ of\ joins}$ & $\bold {0}$ & $\bold {1}$ & $\bold {2}$ & $\bold {3}$ & $\bold {4}$ & $\bold {5}$ & $\bold {overall}$ \\
\hline
$\bold {UnqCrd\_test1}$ & $150$ & $150$ & $150$ & $0$ & $0$ & $0$ & $450$\\
$\bold {UnqCrd\_test2}$ & $75$ & $75$ & $75$ & $75$ & $75$ & $75$ & $450$\\
\hline
\hline
\end{tabular}
}
\caption{Distribution of joins.}
\end{table}

\begin{comment}
All the limited models estimate cardinalities \textit{with} duplicates. We extended these models using the PUNQ model, to estimate cardinality without duplicates. In PostgreSQL, we simply added the DISTINCT keyword in the queries' select clauses, and then used the ANALYZE command to obtain the estimated (set-theoretic) cardinalities.
\end{comment}

\subsection{The Quality of Estimates}
\label{The Quality of Estimates PUNQ}
\begin{figure}[h!]
\begin{center}
  \includegraphics[width=7.8cm]{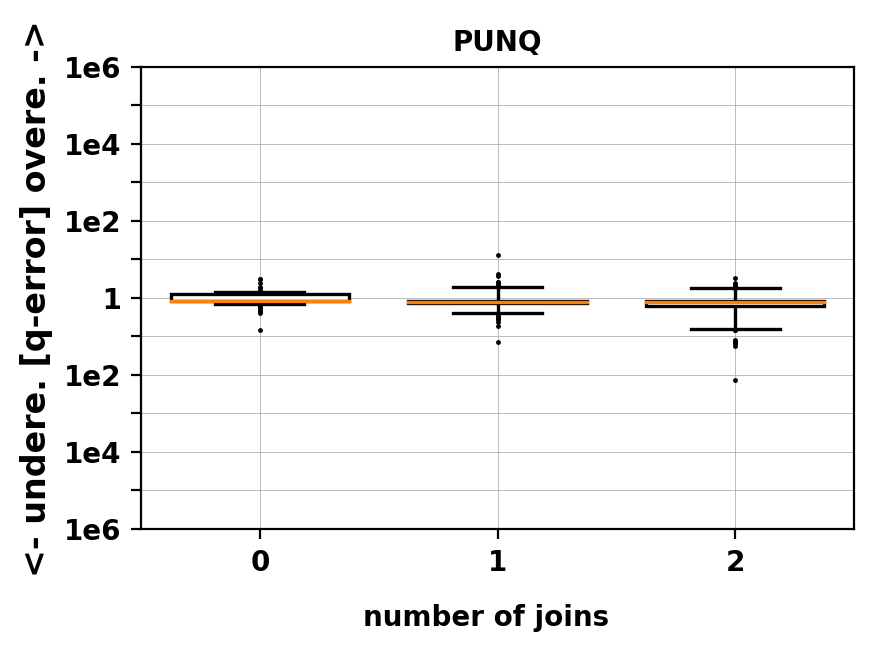}
  \caption{Uniqueness rates estimation q-errors on the UnqCrd\_test1 workload. In all similar presented plots, the box boundaries are at the 25th/75th percentiles and the horizontal lines mark the 5th/95th percentiles. Thus 50\% of the described test results are located in the boxes, and 90\% are located within the horizontal lines (boxes are clearer in subsequent Figures).}
  \label{fig:PUNQ1}
\end{center}
\end{figure}

We examined the UnqCrd\_test1 workload on two state-of-the-art limited models, MSCN and CRN. Recall that this workload includes queries with up to two joins, and that the MSCN, CRN and PUNQ models were trained over such conjunctive queries.
Figure \ref{fig:PUNQ1} and Table \ref{table:PUNQ1} depict the PUNQ model uniqueness rates estimations. It is clear that the PUNQ model estimates are very accurate, where 75\% of the examined queries uniqueness rates were predicted accurately with a q-error smaller than 1.93. That is, the ratio between the estimated rates and the actual rates does not exceed 1.93, whether it is over or under estimated. These results are similar for queries with 0, 1, and 2 joins.

\begin{table}[h!]
\centering
\resizebox{\linewidth}{!}{
\begin{tabular}{lccccccc}
\hline
\hline
$\bold {}$ & $\bold {50th}$ & $\bold {75th}$ & $\bold {90th}$ & $\bold {95th}$ & $\bold {99th}$ & $\bold {max}$ & $\bold {mean}$\\
\hline
$\bold {PUNQ}$ & $1.12$ & $1.93$ & $3.71$ & $5.59$ & $13.65$ & $139$ & $2.27$\\
\hline
\hline
\end{tabular}
}
\caption{Uniqueness rates estimation q-errors on the UnqCrd\_test1 workload. In all similar presented tables, we depict the percentiles, maximum, and mean q-errors of the test (the p’th percentile, is the q-error value below which p\% of the test q-errors are found).}
\label{table:PUNQ1}
\end{table}

The fact that the PUNQ model uniqueness rate estimates ($U$) are very accurate, along with the fact that the CRN and the MSCN models cardinality estimates with duplicates ($C$) are accurate as well, is thus reflected in accurate cardinality estimates without duplicates ($U \cdot C$).

\begin{figure}[h!]
\begin{center}
  \includegraphics[width=\linewidth]{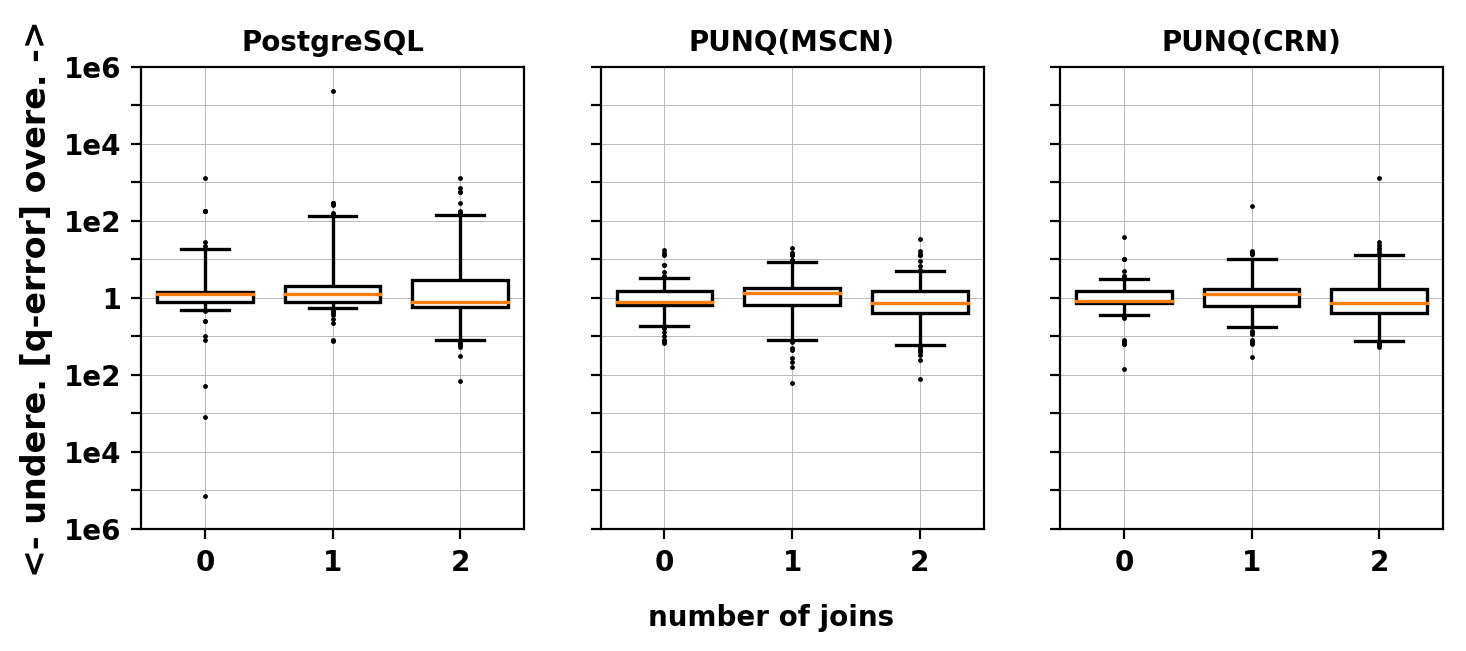}
  \caption{Cardinality estimation q-errors on the UnqCrd\_test1 workload.}
  \label{fig:UnqCrd1}
  \vspace{-2mm}
\end{center}
\end{figure}

As a consequence, examining the results in Figure \ref{fig:UnqCrd1} and Table \ref{table:UnqCrd1}, it is apparent that when the CRN and MSCN models are extended to estimate cardinalities without duplicates using the PUNQ model, the estimates are still very accurate. Observe that the models we extend were originally tailored to estimate cardinalities with duplicates only.
Note that postgreSQL model was not extended using the PUNQ model, as it supports any SQL queries as is. Interestingly, it obtained the worst results among the examined models.

\begin{table}[h!]
\centering
\resizebox{\linewidth}{!}{
\begin{tabular}{lccccccc}
\hline
\hline
$\bold {}$ & $\bold {50th}$ & $\bold {75th}$ & $\bold {90th}$ & $\bold {95th}$ & $\bold {99th}$ & $\bold {max}$ & $\bold {mean}$\\
\hline
$\bold {PostgreSQL}$ & $\bold{1.82}$ & $4.93$ & $27.26$ & $163$ & $926$ & $372207$ & $1214$\\
$\bold{PUNQ(MSCN)}$ & $2.13$ & $4.83$ & $10.05$ & $16.37$ & $60.53$ & $\bold{222}$ & $\bold{5.66}$\\

$\bold {PUNQ(CRN)}$ & $2.05$ & $\bold{4.27}$ & $\bold{9.98}$ & $\bold{15.24}$ & $\bold{49.11}$ & $1037$ & $7.49$\\

\hline
\hline
\end{tabular}
}
\caption{Cardinality estimation q-errors on the UnqCrd\_test1 workload.}
\label{table:UnqCrd1}
\vspace{-2mm}
\end{table}

\subsection{Generalizing to Additional Joins}
\label{Generalizing to Additional Joins PUNQ}
In this section we examine how the PUNQ model generalizes to queries with more than 2 joins, and how limited models adapt to such queries under the PUNQ model extension. To do so we use the UnqCrd\_test2 workload which includes queries with zero to \textit{five} joins. Recall that we trained MSCN, CRN, and PUNQ models with conjunctive queries that have only up to \textit{two} joins.
To examine the limited cardinality estimation models more broadly, we used two additional limited models, the Improved PostgreSQL model, and the Improved MSCN model (both introduced in \cite{CRN}).

\begin{figure}[h!]
\begin{center}
  \includegraphics[width=7.1cm]{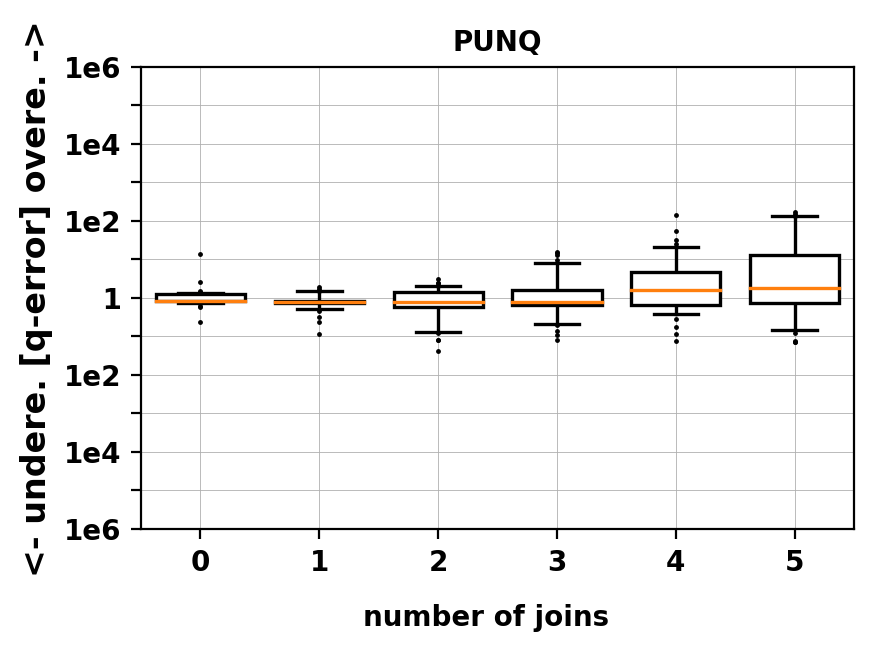}
  \caption{Uniqueness rates estimation q-errors on the UnqCrd\_test2 workload.}
  \label{fig:PUNQ2}
\end{center}
\vspace{-4mm}
\end{figure}

Figure \ref{fig:PUNQ2} and Table \ref{table:PUNQ2} depict the PUNQ model uniqueness rate estimations. Note that the PUNQ model estimates are very accurate when considering queries with zero to two joins, whereas the estimates tend to over-estimate as the number of joins increases. This is to be expected as PUNQ was trained with queries with up to two joins only. Nevertheless, the overall estimates still relatively accurate.

\begin{figure*}[h!]
\begin{center}
  \includegraphics[width=\linewidth]{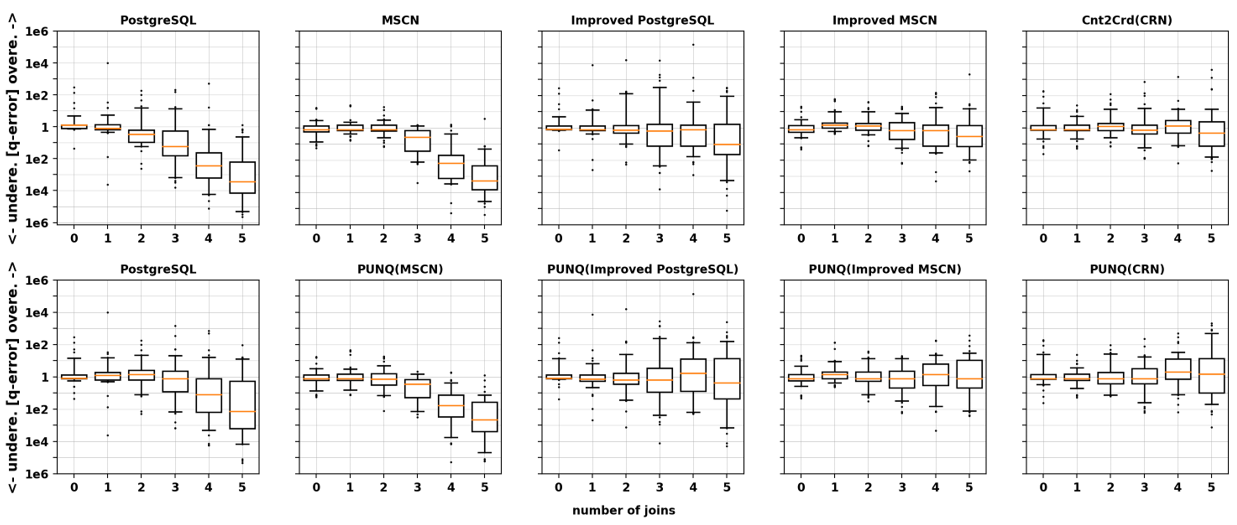}
    \caption{Cardinality estimation q-errors on the UnqCrd\_test2 workload.}
  \label{fig:all}
\end{center}
\end{figure*} 

\begin{table}[h!]
\centering
\resizebox{\linewidth}{!}{
\begin{tabular}{lccccccc}
\hline
\hline
$\bold {}$ & $\bold {50th}$ & $\bold {75th}$ & $\bold {90th}$ & $\bold {95th}$ & $\bold {99th}$ & $\bold {max}$ & $\bold {mean}$\\
\hline
$\bold {PUNQ}$ & $1.75$ & $3.55$ & $9.9$ & $18.08$ & $109$ & $214$ & $6.62$\\
\hline
\hline
\end{tabular}
}
\caption{Uniqueness rates estimation q-errors on the UnqCrd\_test2 workload.}
\label{table:PUNQ2}
\vspace{-4mm}
\end{table}

Figure \ref{fig:all} provides a fuller picture of how the limited models estimates are affected when they are extended to estimate cardinalities without duplicates. In the upper part of Figure \ref{fig:all}, we present the limited models' cardinality estimates errors with duplicates, i.e., the models' original estimates. In the bottom part, we show the models' estimates errors on the same queries, when the models are extended to estimate set-theoretic cardinalities using the PUNQ model. 

According to box-plots definition, 50\% of the test queries are located within the box boundaries, and 90\% are located between the horizontal lines. Observe that the boxes' boundaries and the horizontal lines were hardly changed in the extended models box-plots, compared with the original models box-plots. We conclude that for each model and for each number of joins, the majority of the queries are located in the same error domains. That is, the models' accuracies were hardly changed after extending them using the PUNQ model. They maintained almost the same quality of estimates for set-theoretic cardinality, as the original models' quality for estimating cardinality with duplicates. 

Significant changes in the models quality of estimates may be observed in queries with more than 2 joins. In such queries, the PUNQ model suffers from slightly overestimated uniqueness rates, which directly affect the extended models set-theoretic cardinality estimates. This is depicted via their box-plots horizontal lines moving a bit upwards as compared with the box-plots of the original models estimates.

Table \ref{table:UnqCrd2} shows the cardinality estimates errors percentiles and means using the examined extended models.

\begin{table}[h!]
\centering
\resizebox{\linewidth}{!}{
\begin{tabular}{lccccccc}
\hline
\hline
$\bold {}$ & $\bold {50th}$ & $\bold {75th}$ & $\bold {90th}$ & $\bold {95th}$ & $\bold {99th}$ & $\bold {max}$ & $\bold {mean}$\\
\hline
$\bold {PostgreSQL}$ & $3.95$ & $34.95$ & $651$ & $2993$ & $20695$ & $331152$ & $2029$\\

$\bold{PUNQ(MSCN)}$ & $4.85$ & $59.65$ & $1189$ & $5476$ & $86046$ & $288045$ & $3430$\\

$\bold {PUNQ(Imp. Post.)}$ & $3.51$ & $20.97$ & $134$ & $444$ & $9770$ & $136721$ & $563$\\

$\bold{PUNQ(Imp. MSCN)}$ & $3.15$ & $10.35$ & $33.73$ & $88.61$ & $244$ & $3298$ & $24.46$\\

$\bold{PUNQ(CRN)}$ & $\bold{3.31}$ & $\bold{10.81}$ & $\bold{39.95}$ & $\bold{90.98}$ & $\bold{778}$ & $\bold{3069}$ & $\bold{35.71}$\\

\hline
\hline
\end{tabular}
}
\caption{Cardinality estimation q-errors on the UnqCrd\_test2 workload.}
\label{table:UnqCrd2}
\end{table}

In Table \ref{table:UnqCrd2} and Figure \ref{fig:all}, it is clear that the PUNQ(CRN) and PUNQ(Improved MSCN) models are significantly more robust in generalizing to queries with additional joins as compared with the other examined models. In particular, the two models are more robust than postgreSQL. This is despite the fact that postgreSQL can estimate cardinalities without duplicates by default, without the need to extend it using PUNQ, as was done for the other limited models.

Recall that the extended models' quality of estimates is mainly dependent on the original limited models' quality of estimates. Hence, had we used other more (resp., less) accurate limited  models, we would simply get more (resp., less) accurate results in the extended models. In particular, we can obtain better estimates by extending more accurate models (not necessarily based on ML) that may be developed in the future.

These results highlight the usefulness of the PUNQ model in extending limited models. PUNQ therefore presents a simple tool to extend limited models, while keeping their quality of set-theoretic estimates roughly without change.

\subsection{PUNQ Cardinality Prediction Time}
\label{PUNQ Prediction Time}
As described above, to estimate cardinalities without duplicates by extending limited cardinality estimation models, we used the PUNQ model along with the estimated cardinalities of these models. Therefore, the prediction time of cardinalities without duplicates by extending limited models, consists of two parts. First, the prediction time of the limited model. Second, the prediction time of the uniqueness rate, obtained from the PUNQ model.

The prediction times of the limited models vary, depending on the model itself. Table \ref{table:time1} depicts the average prediction time in milliseconds, for estimating cardinality with duplicates of a single query, using different limited models.

\begin{table}[h!]
\centering
\resizebox{\linewidth}{!}{
\begin{tabular}{lcccc}
\hline
\hline
$\bold {Model}$  & $\bold {MSCN}$ & $\bold {CRN}$ &  $\bold {Imp.\ MSCN}$ & $\bold {Imp.\ PostgreSQL}$ \\
\hline
$\bold {Prediction\ time}$ & 0.25ms &  15ms & 35ms & 75ms  \\
\hline
\hline
\end{tabular}
}
\caption{Average prediction time of a single query.}
\label{table:time1}
\vspace{-1mm}
\end{table}

The prediction time of uniqueness rates using the PUNQ model is 0.05ms, per query (as described in Section \ref{Hyperparameter Search}). Therefore, the set-theoretic cardinality prediction time overhead due to using the PUNQ model, when extending limited models, is minor (0.05ms), as compared with the limited models' prediction times (of cardinalities with duplicates).

\subsection{Supporting Inequality Joins with PUNQ}
Thus far, the PUNQ model described in Section \ref{Model} supported queries that have equality joins only. This was done since most of the recently developed cardinality estimation models $M$ (e.g., MSCN) focus on supporting only such queries. Thus, even if the PUNQ model were able to estimate the uniqueness rates of queries that have inequality joins, still the extended model PUNQ($M$) (e.g., PUNQ(MSCN)) will not be able to estimate set-theoretic cardinalities of such queries (inequality joins queries). This is due to the lack of support of inequality joins in the limited model $M$.

However, supporting inequality joins is interesting and essential in real-world databases. Therefore, we describe a simple modification to the PUNQ model architecture, so that it can support inequality joins as well. As a result, given \textit{any} limited cardinality estimation model $M$ that does support queries that have inequality joins (e.g., PostgreSQL, Improved PostgreSQL or CRN), using the revised PUNQ model, the PUNQ($M$) model is also be able to estimates the set-theoretic cardinalities of such queries.

To support inequality joins in the PUNQ model, we only revised the query vectors' representation at the first stage of the model (Section \ref{First Stage}) as follows (the second and third stages were not changed). We revised the join segments, so that instead of using only the J1-seg and J2-seg segments that were used to represent the (equality) joins, we added a new segment, JO-seg to represent the specific join operator. As a result we can represent inequality joins as well as equality joins. Given any type of join $(col1,op,col2)$, $col1$, $op$ and $col2$ are represented using one-hot vectors placed in the J1-seg, JO-seg and J2-seg segments, respectively. The revised vector segmentation scheme is described in Table \ref{table:Revised Vector Segmentation} (the highlighted cells are the only ones changed).

\begin{table}[h!]
\centering
\resizebox{\linewidth}{!}{
\begin{tabular}{c|c|c|c|c|c|c|c|c}
\hline
\hline
$\bold{Type}$ & $\bold{Att.}$ & $\bold{Table}$ & \multicolumn{3}{c|}{\cellcolor{gray!50} $\bold{Join\ (any\ kind)}$} & \multicolumn{3}{c}{$\bold{Column\ Predicate}$} \\
\hline
$\bold {Segment}$ & \textbf{A-seg} & \textbf{T-seg} & \textbf{J1-seg} & \cellcolor{gray!50} \textbf{JO-seg} & \textbf{J2-seg} &  \textbf{C-seg} & \textbf{O-seg} & \textbf{V-seg} \\
\hline
$\bold {Seg.\ size}$ & $\#C$ & $\#T$ & $\#C$ & \cellcolor{gray!50} $\#O$ &  $\#C$ &  $\#C$ & $\#O$ & $1$ \\
\hline
$\bold {Feat.}$ & $1hot\ vec.$ & $1hot\ vec.$ & $1hot\ vec.$ & \cellcolor{gray!50} $1hot\ vec.$ & $1hot\ vec.$ &  $1hot\ vec.$ & $1hot\ vec.$ & $norm.$ \\
\hline
\hline
\end{tabular}
}
\caption{Revised Vector Segmentation.}
\label{table:Revised Vector Segmentation}
\end{table}

Experimentally, by training the PUNQ model with queries that have inequality joins, we found out the the model convergence time and quality of estimates (of the uniqueness rates) are very similar to those described in the previous sections. In particular, we found out that the number of inequality joins does not have a negative affect to the model estimates as the equality joins do.
Due to limited space, we briefly describe in Table \ref{table:PUNQ'} the q-errors results when examining a workload that includes 720 queries that have inequality joins (equally distributed with zero to \textit{five} equality joins).

\begin{table}[h!]
\centering
\resizebox{\linewidth}{!}{
\begin{tabular}{lccccccc}
\hline
\hline
$\bold {}$ & $\bold {50th}$ & $\bold {75th}$ & $\bold {90th}$ & $\bold {95th}$ & $\bold {99th}$ & $\bold {max}$ & $\bold {mean}$\\
\hline
$\bold {PUNQ}$ & $1.79$ & $5.29$ & $16.35$ & $33.84$ & $130$ & $274$ & $9.53$\\
\hline

$\bold {PostgreSQL}$ & $4.94$ & $21.18$ & $185$ & $4417$ & $25832$ & $487069$ & $2411$\\

$\bold {PUNQ(Imp. Post.)}$ & $3.18$ & $17.26$ & $123$ & $414$ & $4932$ & $44407$ & $307$\\

$\bold{PUNQ(CRN)}$ & $2.94$ & $8.96$ & $47.72$ & $149$ & $1175$ & $7804$ & $47.94$\\

\hline
\hline
\end{tabular}
}
\caption{Q-errors on inequality joins queries workload.}
\label{table:PUNQ'}
\vspace{-1mm}
\end{table}

\section{Supporting AND, OR \& NOT}
\label{Supporting AND, OR, NOT operators}
Several cardinality estimation models consider only conjunctive queries (i.e., queries that only use the AND operator). Other frequently used operators are the OR and NOT operators.
To estimate the cardinalities of queries with OR and NOT with such models, we need to change their architectures. Changing architectures is complex, as each model is structured in a different way.
Therefore, we introduce a uniform alternative approach for estimating cardinalities for general queries, namely the GenCrd algorithm.
Using this approach, we can use \textit{any} limited model, that only supports conjunctive queries, to estimate cardinalities for queries with the AND, OR, and NOT operators.

\subsection{The GenCrd Algorithm}
\label{GenCrd}
Consider estimating the cardinality of a general query $Q$. The GenCrd algorithm relies mainly on two observations.

\subsubsection{First Observation}
Given a general query $Q$ that includes AND, OR, and NOT operators, we can represent $Q$ as multiple conjunctive queries, union-ed with OR. That is, query $Q$ can be transformed to a query of the form $Q_1\ OR\ Q_2\ OR\ ...\ OR\ Q_n$. We can therefore represent query $Q$ as a list of conjunctive queries $[Q_1,Q_2,...,Q_n]$ where each $Q_i$ includes only AND operators (with the same SELECT and FROM clauses as in query $Q$). This is done by converting $Q$'s WHERE clause into disjunctive normal form (DNF) \cite{DNF1,DNF2,DNF3}, using simple logical transformation rules, and by considering each conjunctive disjunct as a separate query. For simplicity, we refer to this list $[Q_1,Q_2,...,Q_n]$ as the \textit{DNF-list}, and denote it as $Q_{1,2,...,n}$.

\subsubsection{Second Observation}
Consider estimating the cardinality of a general query $Q$, using its representing \textit{DNF-list} of conjunctive queries  $[Q_1,Q_2,...,Q_n]$. Query $Q$'s cardinality can be calculated by a simple algorithm, as follows\footnote{We use $| \cdot |$ but we could also have used $|| \cdot ||$ ($| \cdot |$ and $|| \cdot ||$ are defined in Section \ref{Uniqueness Rate Definition}).}:
\begin{itemize}
    \item Calculate $a = |Q_1|$.
    \item Calculate $b = |Q_{2,3,...,n}|$.
    \item Calculate $c = |Q_1\ \intersection\ Q_{2,3,...,n}|$.
    \item Then, $|Q| = a + b - c.$
\end{itemize}
Quantity $a$ can be calculated using any cardinality estimation model that supports conjunctive queries. Quantity $b$ is calculated, recursively, using the same algorithm, since the list contains only conjunctive queries, and forms a proper input for the algorithm.
Similarly, quantity $c$ is calculated, recursively, as described below:
\begin{equation}
\label{exp1}
  |Q_1\ \intersection\ Q_{2,3,...,n}| = |\ Q_1\ \intersection\ [Q_2,Q_3,...,Q_n]\ |
\end{equation}
Note that Equation \ref{exp1} is equivalent to Equation \ref{exp2}:
\begin{equation}
\label{exp2}
|\ [Q_1\ \intersection\ Q_2,Q_1\ \intersection\ Q_3,...,Q_1\ \intersection\ Q_n]\ |
\end{equation}
Therefore, we use the same recursive algorithm since the resulting list also contains conjunctive queries and forms a proper input with fewer queries.

The algorithm is exponential in the size of the \textit{DNF-list}. Given a query $Q$ for cardinality estimation, using the GenCrd algorithm, we call the cardinality estimation model at most C(m) times, where m is the size of the representing \textit{DNF-list}:
$$ C(m) = 2^m -1 $$
Note that C(m) is an upper bound. As the number of OR operators is usually small, this expression is practically not prohibitive.
Additionally, in Section \ref{The ImplyFalse Algorithm} we describe the ImplyFalse algorithm that reduces the potential exponential number of calls to the limited cardinality estimation model to practically linear number.
\begin{figure}[!h]
    \begin{mdframed}
    GenCrdRec($DNF\_list$):\\
    \hspace*{0.3cm} if len($DNF\_list$) == 1:\\
    \hspace*{0.6cm} if ImplyFalse($DNF\_list$[0]):\\
    \hspace*{0.9cm} return 0\\
    \hspace*{0.6cm} return Cardinality($DNF\_list$[0])\\
    \hspace*{0.3cm} else:\\
    \hspace*{0.6cm} $cnj_\_q$ = [$DNF\_list$[0]] \\
    \hspace*{0.6cm} $smaller\_list$ = $DNF\_list$[1:end] \\
    \hspace*{0.6cm} $updated\_list$ = [$q\  \intersection\ cnj\_q$ For $q$ in  $smaller\_list$ ]\\ \\
    \hspace*{0.6cm} return GenCrdRec($cnj\_q$)\\
    \hspace*{1.5cm} + GenCrdRec($smaller\_list$)\\
    \hspace*{1.5cm} - GenCrdRec($updated\_list$)\\ \\
    GenCrd($Q$):\\
    \hspace*{0.3cm} return GenCrdRec(GetDNFlist($Q$))
    \end{mdframed}
    \caption{The GenCrd Algorithm.}
    \label{GenCrd Algorithm}
\end{figure}

\begin{figure*}
  \includegraphics[width=\linewidth]{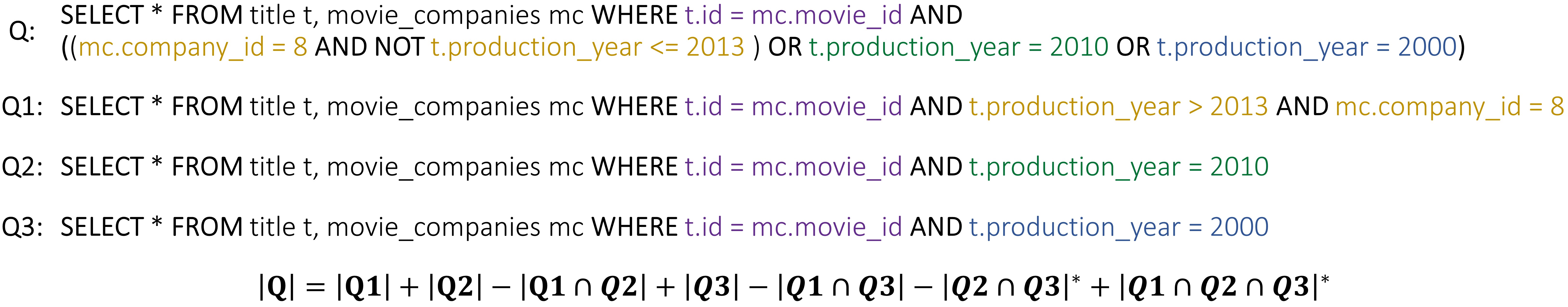}
  \caption{Converting a general SQL query \textit{Q} with AND, OR, and NOT operators into a DNF form, creating a list of conjunctive queries [Q1, Q2, Q3]. \textit{Q}'s cardinality is then calculated using the GenCrd algorithm with the \textit{DNF-list} as input. The starred queries imply False, hence, they have zero-cardinality and are detected using the ImplyFalse algorithm.}
  \label{fig:algo}
\end{figure*}

In Figure \ref{GenCrd Algorithm}, $Q1 \intersection Q2$ is the intersection query of $Q1$ and $Q2$ whose SELECT and FROM clauses are identical to $Q1$’s (and $Q2$’s) clauses, and whose WHERE clause is $Q1$’s AND $Q2$’s WHERE clauses. Function GetDNFlist returns the list of conjunctive queries representing query $Q$. Function Cardinality($Q$) can be implemented by using \textit{any} limited cardinality estimation model for estimating the cardinality of the given input conjunctive query $Q$ (see a simple example in Figure \ref{fig:algo}).

As described in Figure \ref{GenCrd Algorithm}, during the execution of the GenCrd algorithm we create multiple conjunctive queries from multiple smaller queries. These queries may often contain contradictory predicates. Therefore, to reduce the prediction time and errors, we use the ImplyFalse algorithm, before directly using the cardinality estimation models.
If ImplyFlase returns True, then query $Q$ has zero-cardinality, and therefore we do not need to call the cardinality estimation model subroutine. This way, the actual number of times we call the cardinality estimation model is practically smaller than the upper bound given in the formula for C(m).

Depending on its implementation, Function Cardinality($Q$) returns the cardinality of $Q$ with, or without, duplicates. In the first (resp., second) case, GenCrd will therefore return cardinalities with (resp., without) duplicates.  We exhibited a generic method to estimate set-theoretic cardinalities, using PUNQ. Thus, if needed, we can implement function Cardinality($Q$) such that it returns set-theoretic cardinalities.

\newpage
\subsection{The ImplyFalse Algorithm}
\label{The ImplyFalse Algorithm}
The ImplyFalse algorithm takes as input a conjunctive query ($Q$) with equality joins, and checks whether there are any contradictory predicates in $Q$.

The ImplyFalse algorithm runs in four main stages, as described in Figure \ref{ImplyFalse Algorithm}.
(1) It first initializes three maps with single element classes. Initially, each class includes a single column, with initial values accordingly. (2) In the first loop, it unions the classes of the columns that must be equal using the function UnionClasses(c1,c2) which unions the classes of columns c1 and c2 into a single class. Hence, each class includes all the columns that must have equal values. (3) In the second loop, it updates the maps according to the columns' predicates. (4) Finally, in the last loop, it checks whether there are any \textit{contradictory} predicates.

Determining whether a conjunctive Boolean expression is equivalent to False has been shown to be a co-NP-complete problem in its full generality \cite{ulmanBook,foundationBook}. 
However, our case is tractable, and the problem is solved in linear time in the number of columns used in the input query, as described in Figure \ref{ImplyFalse Algorithm}. 
This is due to the form of the examined conjunctive queries (comparison to constants). The examined conjunctive queries include joins of one type only, equality join (col1,op,col2) where op is =. In addition, the columns' predicates are of the form (col,op,val) s.t. op $\in [<,=,>]$ \footnote{Operator $\leq$ can be expressed with $<,=$ and OR. Similarly, operator $\geq$.}.

\begin{figure}[!h]
    \begin{mdframed}
    ImplyFalse(Query $Q$):\\
    \hspace*{0.3cm} minVals = \{ [col]: $-\infty$ $|$ col is a column used in $Q$\}\\
    \hspace*{0.3cm} maxVals =  \{ [col]: $\infty$ $|$ col is a column used in $Q$\}\\
    \hspace*{0.3cm} exactVals = \{ [col]: $\bot$ $|$ col is a column used in $Q$\}\\ \\
    \hspace*{0.3cm} For-each join clause (col1,$=$,col2) in $Q$:  \\  
    \hspace*{0.6cm} UnionClasses(col1,col2)\\ \\
    \hspace*{0.3cm} For-each column predicate (col,op,val) in $Q$: \\
    \hspace*{0.6cm} if op $==$ '$>$':\\
    \hspace*{1.2cm} minVals[col] $=$ Max(val, minVals[col])\\  
    \hspace*{0.6cm} if op $==$ '$<$':\\
    \hspace*{0.9cm} maxVals[col] $=$ Min(val, maxVals[col]) \\ 
    \hspace*{0.6cm} if op $==$ '$=$':\\
    \hspace*{0.9cm} if exactVals[col] $\neq \bot$ \textit{and} exactVals[col] $\neq$ val:\\
    \hspace*{1.2cm} return True\\
    \hspace*{0.9cm} exactVals[col] $=$ val\\ \\
    \hspace*{0.3cm} For-each col in all the columns used in $Q$: \\
    \hspace*{0.6cm} if maxVals[col] $\leq$ minVals[col]:\\
    \hspace*{0.9cm} return True\\
    \hspace*{0.6cm} if exactVals[col] $\neq \bot$ \textit{and not} \\ 
    \hspace*{0.75cm} minVals[col] $\leq$ exactVals[col] $\leq$ maxVals[col]:\\
    \hspace*{0.9cm} return True \\ \\
    \hspace*{0.3cm} return False
    \end{mdframed}
    \caption{The ImplyFalse Algorithm.}
    \label{ImplyFalse Algorithm}
\end{figure}

In Figure \ref{ImplyFalse Algorithm}, the Map[$c$] operator returns the corresponding value of the appropriate class in which column $c$ is located. The $\infty, \bot$ symbols denote an infinity and uninitialized value, respectively. Functions Min(x,y) and Max(x,y) return the minimum and the maximum between x and y, respectively.

The end result of introducing the GenCrd algorithm is efficiently and effectively extending estimation capabilities to a vaster class of queries. This \textit{eliminates} the need to train models from scratch on this gigantic class of queries. It also factors out this code for \textit{any} optimizer which results in a better software architecture.

\section{GenCrd Evaluation}
\label{GenCrd Evaluation}
In this section, we examine how well the GenCrd algorithm, transforms limited cardinality estimation models that only handles conjunctive queries into ones that support queries with the AND, OR, and NOT operators. To do so, we extend the same four models that were used in the PUNQ evaluation, as described in Section \ref{PUNQ Evaluation}. Each model was extended by simply using the model in the GenCrd algorithm as a subroutine, replacing the Cardinality function used in the algorithm, see Figure \ref{GenCrd Algorithm}.
For clarity, we denote the GenCrd extended model of $M$ as GenCrd($M$).

We extend and evaluate each of the four examined limited models over two test workloads. In addition, we compare our results with those of the PostgreSQL version 11 cardinality estimation component \cite{postgreSQL}, as postgreSQL supports all SQL queries. To estimate cardinality of the examined queries using PostgreSQL, we use the ANALYZE command to obtain the estimated cardinalities. 

The workloads were created using the same queries generator (using a different random seed) as presented in Section \ref{queries generator} with an additional step. Since the queries generator creates only conjunctive queries, we reconfigured it as follows. After generating a conjunctive query $Q$, it randomly chooses a set $P$ out of $Q$'s column predicates. For each column predicate $p=(col,op,val)$ in the chosen set $P$, it randomly creates a new predicate $p'=(col',op',val')$ and replaces the original column predicate $p$ in $Q$ with the predicate $p\ OR\ p'$.
$col'$ may be the same as $col$ or changed randomly. This also holds for $op'$ and $val'$. We make sure that at least one of them is changed so that $p'$ is not equivalent to $p$.
Subsequently, the queries generator chooses another set $P$ from the updated $Q$'s column predicates. Each column predicate $p$, in the chosen set $P$, is replaced with the predicate $NOT\ p$ in $Q$. This way, we obtain a query that includes the AND, OR, and NOT operators.

In order to ensure a fair comparison, the CRN and MSCN models were trained similarly, as noted in section \ref{PUNQ Evaluation}.

\newpage
\subsection{Evaluation Workloads}
\label{GenCrd Evaluation Workloads}
The evaluation uses the IMDb dataset, over two different query workloads:
\begin{itemize}
    \item GenCrd\_test1, a synthetic workload with 450 unique queries, with zero to two joins.
    \item GenCrd\_test2, a synthetic workload with 450 unique queries, with zero to \textit{five} joins. This dataset is designed to examine how the models generalize beyond 2 joins.
\end{itemize}

Both workloads are equally distributed in the number of joins. From these workloads we generated \textit{DNF-lists}, each representing a general query. For each number of joins, the queries are uniformly distributed in terms of the size of the representing \textit{DNF-list}, from 1 to 5. That is, each query has representing \textit{DNF-lists} of sizes 1 (conjunctive query without any OR and NOT operators), up to 5 (general query whose representing \textit{DNF-list} includes 5 such conjunctive queries).

\begin{table}[h!]
\centering
\resizebox{\linewidth}{!}{
\begin{tabular}{lccccccc}
\hline
\hline
$\bold {number\ of\ joins}$ & $\bold {0}$ & $\bold {1}$ & $\bold {2}$ & $\bold {3}$ & $\bold {4}$ & $\bold {5}$ & $\bold {overall}$ \\
\hline
$\bold {GenCrd\_test1}$ & $150$ & $150$ & $150$ & $0$ & $0$ & $0$ & $450$\\
$\bold {GenCrd\_test2}$ & $75$ & $75$ & $75$ & $75$ & $75$ & $75$ & $450$\\
\hline
\hline
\end{tabular}
}
\caption{Distribution of joins. For each number of joins the representing \textit{DNF-list} size is equally distributed from 1 to 5.}
\end{table}

\begin{comment}
In the experiments described in Sections \ref{The Quality of Estimates} and \ref{Generalizing to Additional Joins}, all the examined models, except for the PostgreSQL model, are limited to supporting only conjunctive queries. Therefore, we used the GenCrd algorithm with these models embedded within it to estimate the cardinalities of the examined queries. In PostgreSQL, we used the ANALYZE command to obtain the estimated cardinalities.
\end{comment}

\subsection{The Quality of Estimates}
\label{The Quality of Estimates}
We examined the GenCrd\_test1 workload using two state-of-the-art limited models, MSCN and CRN. Recall that this workload includes queries with up to two joins, as MSCN and CRN were trained over such conjunctive queries. 

\begin{figure}[h!]
\begin{center}
  \includegraphics[width=\linewidth]{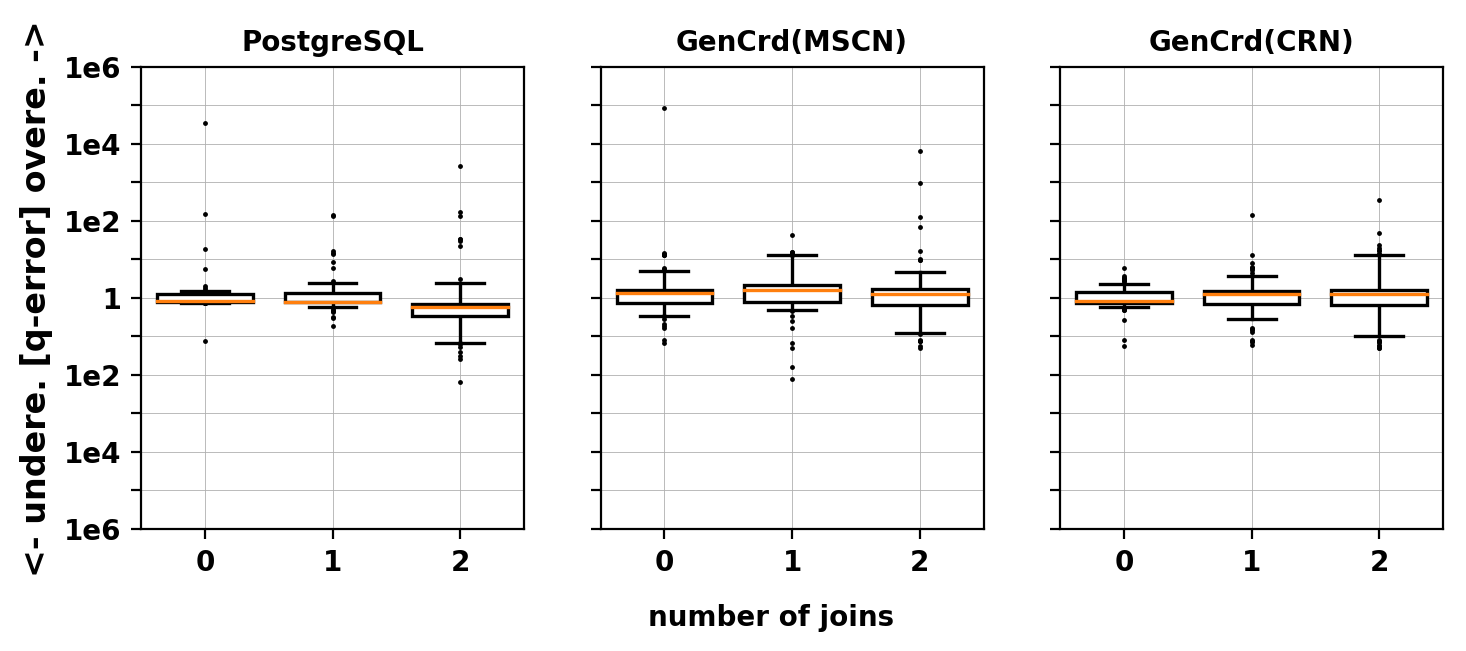}
  \caption{Cardinality estimation q-errors on the GenCrd\_test1 workload.}
  \label{fig:GenCrd1}
\end{center}
\end{figure}

Although the MSCN and CRN models were initially tailored to estimate cardinalities for conjunctive queries only, examining the results in Figure \ref{fig:GenCrd1} and Table \ref{table:GenCrd1}, it is apparent that these models are successfully extended to estimate cardinalities accuratly for general queries using the GenCrd algorithm.

\begin{figure*}[h!]
\begin{center}
  \includegraphics[width=\linewidth]{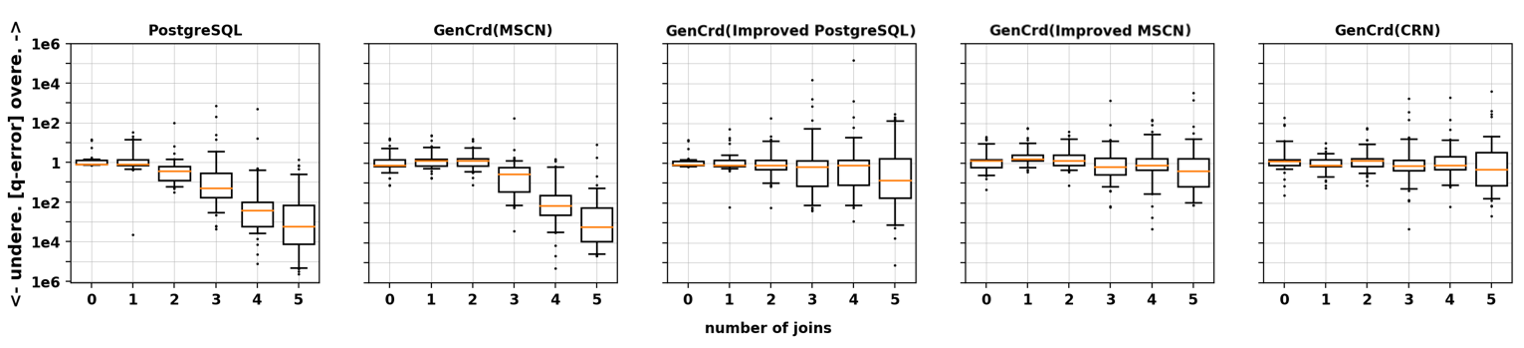}
    \caption{Cardinality estimation q-errors on the GenCrd\_test2 workload.}
  \label{fig:GenCrd2}
\end{center}
\vspace{-1mm}
\end{figure*}

\begin{table}[h!]
\centering
\resizebox{\linewidth}{!}{
\begin{tabular}{lccccccc}
\hline
\hline
$\bold {}$ & $\bold {50th}$ & $\bold {75th}$ & $\bold {90th}$ & $\bold {95th}$ & $\bold {99th}$ & $\bold {max}$ & $\bold {mean}$\\
\hline
$\bold {PostgreSQL}$ & $\bold{1.32}$ & $2.57$ & $7.42$ & $16.57$ & $154$ & $52438$ & $132$\\
$\bold{GenCrd(MSCN)}$ & $2.01$ & $3.59$ & $7.68$ & $11.97$ & $94.12$ & $92684$ & $232$\\
$\bold {GenCrd(CRN)}$ & $1.68$ & $\bold{2.83}$ & $\bold{6.45}$ & $\bold{10.65}$ & $\bold{30.43}$ & $\bold{538}$ & $\bold{4.83}$\\
\hline
\hline
\end{tabular}
}
\caption{Cardinality estimation q-errors on the GenCrd\_test1 workload.}
\label{table:GenCrd1}
\end{table}

Note that the GenCrd algorithm is an "analytic" algorithm, therefore it does not include any training, in contrast of the PUNQ model. Therefore, when using the GenCrd algorithm, the number of joins, in queries in the DNF-list, has no direct effect on the results accuracy. The quality of estimates merely depends on the quality of the limited cardinality estimation models (that are used as subroutines in the GenCrd algorithm).

\begin{comment}
In contrast, the PUNQ model estimations, are intrinsically affected by the number of joins, and it over-estimates as the number of joins increases.
\end{comment}

\subsection{Generalizing to Additional Joins}
\label{Generalizing to Additional Joins}
Here we examine the GenCrd algorithm with queries with more than 2 joins. To do so we use the GenCrd\_test2 workload which includes queries with zero to \textit{five} joins. 

As can be seen in Figure \ref{fig:GenCrd2}, queries with 3 joins and more have poorer estimates, as compared with the estimates of queries with zero to two joins. This decline in quality is not directly due to the GenCrd algorithm. The decline stems from training the MSCN and CRN models over conjunctive queries that have only up to two joins. Thus, when MSCN and CRN are used as GenCrd subroutines, DNF-list  queries with more joins are estimated not as well by the original MSCN, or CRN, models. 

Comparing conjunctive queries' cardinality estimates using a limited model ($M$), and general queries' cardinality estimates using the extended model (GenCrd($M$)), it appears that there is no significant change in overall quality in terms of q-error. That is, GenCrd maintains the same quality of estimates as the original model's quality of estimates, when $M$ is used as subroutine. This is expected, since the GenCrd($M$) algorithm obtains an estimation for a given query $Q$ by simply summing and subtracting several cardinality estimates (see example in Figure \ref{fig:algo}), where each of these estimates is obtained using the limited model ($M$).

\begin{comment}

To emphasize this point, we provide in table \ref{GenCrd2DIFF} the mean and median q-errors for each of model $M$, and the extended model GenCrd($M$). By comparing these values, it's clear that there is no significant change in the models quality. That is, GenCrd maintain the same quality of estimates as the original models that are used as subroutines.

\begin{table}[h!]
\centering
\resizebox{\linewidth}{!}{
\begin{tabular}{lcccc}
\hline
\hline
$\bold {Model\ M}$ & $\bold {MSCN}$ & $\bold {Imp. post.}$ & $\bold {Imp. MSCN.}$ & $\bold {CRN}$\\
\hline
$\bold {M\ mean}$ & $2940$ & $663$ & $28.73$ & $40.72$ \\
$\bold {GenCrd(M)\ mean}$ & $2611$ & $750$ & $38.43$ & $47.24$ \\
\hline
$\bold {M\ median}$ & $0$ & $0$ & $0$ & $0$ \\
$\bold {GenCrd(M)\ median}$ & $4.17$ & $2.18$ & $2.89$ & $2.26$ \\
\hline
\hline
\end{tabular}
}
\caption{Estimation means and medians.}
\label{table:GenCrd2}
\end{table}

\end{comment}

Table \ref{table:GenCrd2} and Figure \ref{fig:GenCrd2} display the experimental results over the GenCrd\_test2 workload. 
Recall that all the limited models were initially tailored to estimate cardinalities for conjunctive queries only. Hence, these results highlight the usefulness of GenCrd method in extending limited models. 

\begin{table}[h!]
\centering
\resizebox{\linewidth}{!}{
\begin{tabular}{lccccccc}
\hline
\hline
$\bold {}$ & $\bold {50th}$ & $\bold {75th}$ & $\bold {90th}$ & $\bold {95th}$ & $\bold {99th}$ & $\bold {max}$ & $\bold {mean}$\\
\hline
$\bold {PostgreSQL}$ & $8.57$ & $168$ & $3139$ & $12378$ & $316826$ & $647815$ & $8811$ \\
$\bold{GenCrd(MSCN)}$ & $4.17$ & $84.92$ & $1887$ & $6769$ & $60405$ & $278050$ & $2611$\\
$\bold {GenCrd(Impr.\ Post.)}$ & $2.18$ & $10.97$ & $82.75$ & $286$ & $2363$ & $162894$ & $750$\\
$\bold{GenCrd(Impr.\ MSCN)}$ & $2.89$ & $8.45$ & $27.1$ & $73.59$ & $537$ & $5183$ & $38.43$\\
$\bold {GenCrd(CRN)}$ & $\bold{2.26}$ & $\bold{6.03}$ & $\bold{17.49}$ & $\bold{71.17}$ & $\bold{632}$ & $\bold{6025}$ & $\bold{47.24}$\\
\hline
\hline
\end{tabular}
}
\caption{Cardinality estimation q-errors on the GenCrd\_test2 workload.}
\label{table:GenCrd2}
\end{table}

\subsection{GenCrd Cardinality Prediction Time}
\label{GenCrd Prediction Time}
Examining the results of the experiments in Sections \ref{The Quality of Estimates} and \ref{Generalizing to Additional Joins}, we find that the size of the \textit{DNF-list} does not affect the quality of the estimates. That is, queries with representing \textit{DNF-lists} of different sizes, are estimated similarly as long as they have the same number of joins (the higher the number of joins is, the worst the results are). However, the \textit{DNF-list} size directly affects the prediction time. The larger the \textit{DNF-list} is, the larger the prediction time is.

Table \ref{table:time2} depicts the average prediction time in milliseconds, for estimating the cardinality of a single query, when examining different models, with different \textit{DNF-list} sizes.

Note that the PostgreSQL prediction time is not affected by the size of the \textit{DNF-list}, since it does not use GenCrd. 

\begin{table}[h!]
\centering
\resizebox{\linewidth}{!}{
\begin{tabular}{lccccc}
\hline
\hline
$\bold {Model\ \diagdown\ DNF-list\ size}$ & $\bold {1}$ & $\bold {2}$ & $\bold {3}$ & $\bold {4}$ & $\bold {5}$ \\
\hline
$\bold {PostgreSQL}$ & 1.82 & 1.82 & 1.82 & 1.82 & 1.82 \\
$\bold {GenCrd(MSCN)}$ & 0.25 & 0.6 & 0.87 & 1.25 & 1.78 \\
$\bold {GenCrd(Imp.\ Post.)}$ & 75 & 166 & 261 & 374 & 541 \\
$\bold{GenCrd(Imp.\ MSCN)}$ & 35 & 77 & 120 & 186 & 254\\
$\bold{GenCrd(CRN)}$ & 15 & 33 & 53 & 72 & 105 \\
\hline
\hline
\end{tabular}
}
\caption{Avg prediction time of a single query in ms.}
\label{table:time2}
\end{table}

Despite the prediction time increase, time is still in the order of milliseconds \footnote{On average, over the GenCrd\_test2 workload's queries, the prediction time for a single query when using GenCrd(CRN), is 55 ms, while the actual query's execution time is 240000 ms (4 minutes).}. Furthermore, the GenCrd algorithm can easily be parallelized by estimating the cardinality of all the sub-queries in parallel. I.e., by parallelizing all the calls made to the limited model. The estimation times described in Table \ref{table:time2} are based on the sequential version of GenCrd.

\section{Related Work}
\label{Related Work}
Conjunctive queries have been intensely researched; see for example  \cite{ulmanBook,foundationBook,dependencies}. Queries that include the AND, OR, and NOT operators in their WHERE clauses constitute a broad class of frequently used queries. Their expressive power is roughly equivalent to that of the relational algebra. Therefore, this class of queries had been extensively researched early on by the DB theory community. Yannakakis and Sagiv \cite{equ2} showed that testing equivalence of relational expressions with the operators select, project, join, and union is complete for the $\Pi_2^p$ of the polynomial-time hierarchy. Chandra and Merlin \cite{ChandraMerlin} showed that determining containment of conjunctive queries is an NP-complete problem. This also holds under additional settings \cite{ulmanBook,foundationBook,dependencies}.

Estimating cardinalities of such queries was also intensely researched early on, due to its implications for query optimization \cite{SystemR}. Many techniques were proposed to solve this problem, e.g., Random Sampling techniques \cite{RS1,RS2}, Index based Sampling \cite{IBJS}, and recently using neural networks (NN) \cite{CRN,MSCN}.

The introduction of machine learning techniques led to significant improvements in many known estimation problems in databases. Query indexing \cite{MLindex}, query optimization \cite{MLoptimiztion1,MLoptimiztion2}, concurrency control \cite{concurrency}, are some of the problems for which solutions were improved using machine learning techniques. In particular, cardinality estimates for conjunctive queries were significantly improved, by using two recent NN-based models, MSCN \cite{MSCN}, and the CRN models \cite{CRN}. 

The MSCN model \cite{MSCN} is a sophisticated NN tailored for representing conjunctive SQL queries and predicting cardinalities. Technically, it ingeniously presents queries that vary in their structure to a single fixed NN. MSCN has been shown to be superior in estimating cardinalities for conjunctive queries that have the same number of joins as that in the queries training dataset. However, MSCN proved less effective when considering queries with more than two joins as it is not trained on such queries. 

Extending \cite{MSCN}, the work \cite{CRN} proposed a method, Cnt2Crd, for estimating cardinalities using containment rates. Cnt2Crd requires a queries pool that maintains information about previously executed queries. Given a new query $Q$, whose cardinality needs to be estimated, the Cnt2Crd method first estimates the containment rates of the input query $Q$ with the other queries in the pool. This is done using any containment rate estimation model. (\cite{CRN} uses a NN-based model for this task). Then, using the true cardinalities of the queries saved in the queries pool, along with associated containment rates, $Q$'s estimated cardinality is obtained. This approach provides a higher estimation quality for estimating cardinalities of conjunctive queries, compared with other models (e.g. MSCN), especially, in case there are multiple joins. 

Recently, a specialized model for estimating set-theoretic cardinalities was proposed in \cite{AIDB2019}. The model proposed in \cite{AIDB2019} is very similar to the MSCN model, with minor changes in the model architecture. However, the \cite{AIDB2019} model estimates set-theoretic cardinalities on \textit{single} tables only (queries with 0 joins). In this paper, we proposed an alternative approach, PUNQ, that can be adapted to \textit{any} model, and in particular to the MSCN model, without the need to change it. As a result, PUNQ(MSCN) supports a larger and harder class of queries (which include several joins than supported by \cite{AIDB2019}).

\section{Conclusions and Future Work}
\label{Conclusion}
Estimating set-theoretic cardinalities (i.e., without duplicates), is a fundamental problem in query planning and optimization.
Recently suggested NN-based models only handle the case of predicting cardinality with duplicates.
Converting each such model to provide set-theoretic estimates is complex.
We introduce a uniform way for conversion by introducing a neural network, PUNQ, for providing the ratio between the set-theoretic estimate and the estimate with duplicates.

Another deficiency of recent models is the lack of support for more general queries, i.e., ones employing the AND, OR and NOT operators.
To overcome this deficiency without altering existing restricted models (that handle only conjunctive queries), we introduce a recursive algorithm, GenCrd. GenCrd provides estimates for queries employing the AND, OR and NOT operators. GenCrd uses existing models, unaltered, as subroutines. This provides a uniform extension of any restricted estimator into one supporting general queries.

For both extensions, providing set-theoretic estimates and estimating richer queries, we conducted extensive experimentation. The experiments show that both extensions are accurate and highly practical. Thus, we believe that these extension methods are highly promising and practical for solving the cardinality estimation problem.

Thus far, for the PUNQ model, we assumed that the database is static, i.e., no updates. The training of the PUNQ model was performed on an immutable snapshot of the database. Nevertheless, changes in the database content and even in the database schema will occur in a real-world database. Upon schema change, we can rebuild the model according to the updated schema, and retrain it, with a new queries training set; recall that the PUNQ model's training time is in tens of minutes. For retraining the PUNQ model, we need to include the compute cost for obtaining the updated uniqueness rates of queries (on the updated database). 

More practically, we can \textit{incrementally} train the PUNQ model, with new updated queries along with their uniqueness rates. In this solution, handling updates to the database schema is difficult as the PUNQ model neural networks’ dimensions are determined at the outset, according to the number of tables and columns in the schema when the PUNQ model is first built. To solve this, when first building the PUNQ model we can use larger numbers than the actual numbers of tables and columns. This allows adding tables and columns easily and performing retraining incrementally.

GenCrd is an analytic algorithm, which is not based on a specific database.
Thus, unlike the PUNQ model, the GenCrd algorithm is not directly affected by changes that occur in the database. However, the models that are used in it as subroutines (e.g., MSCN or CRN), are affected by changes in the database. Therefore, in order to ensure accurate estimates when using the GenCrd method, the models that are used as subroutines in GenCrd need be up-to-date.

Interestingly, with the extensions we presented, we can estimate cardinalities of queries with GROUP BY (the cardinality is that of the same query with a modified SELECT CLAUSE, i.e., selecting on DISTINCT grouping attributes).  In the future, we plan to generalize our extensions to queries with a HAVING clause as well as ones using EXISTS.

\newpage
\balance
\bibliographystyle{abbrv}
\bibliography{mybib}

\end{document}